\documentclass[times, 10pt,twocolumn,pdftex]{article} 
\usepackage{latex8}
\usepackage{times}
\usepackage{graphicx}
\usepackage{amsmath,amssymb}

\def\@begintheorem#1#2{\list{}{\thm@body}%
  \item[]{\bf #1~#2.}\quad\it\ignorespaces}
\def\@opargbegintheorem#1#2#3{\list{}{\thm@body}%
  \item[]{\bf #1~#2~\ifrembrks #3\global\rembrksfalse\else (#3)\fi.}%
  \quad\it\ignorespaces}
\def\@endtheorem{\endlist}

\newtheorem{definition}{Definition}

\begin{document}

\title{Demand forecasting for companies with many branches, low sales numbers per product, and non-recurring orderings}

\author{Sascha Kurz and J\"org Rambau\\
University of Bayreuth\\ Departement of Mathematics \\ 95440 Bayreuth, Germany\\ \{sascha.kurz,j\"org.rambau\}@uni-bayreuth.de\\
}

\maketitle
\thispagestyle{empty}

\begin{abstract}
  We propose the new Top-Dog-Index to quantify the historic deviation of the
  supply data of many small branches for a commodity group from sales data. On
  the one hand, the common parametric assumptions on the customer demand
  distribution in the literature could not at all be supported in our
  real-world data set.  On the other hand, a reasonably-looking non-parametric
  approach to estimate the demand distribution for the different branches
  directly from the sales distribution could only provide us with
  statistically weak and unreliable estimates for the future demand.
  Based on real-world sales data from our industry partner we provide evidence
  that our Top-Dog-Index is statistically robust.  Using the Top-Dog-Index, we
  propose a heuristics to improve the branch-dependent proportion between
  supply and demand.  Our approach cannot estimate the branch-dependent demand
  directly. It can, however, classify the branches into a given number of
  clusters according to an historic oversupply or undersupply.  This
  classification of branches can iteratively be used to adapt the branch
  distribution of supply and demand in the future.
\end{abstract}

\Section{Introduction}

Many retailers have to deal in their daily businesses with small profit
margins.  Their economic success lies mostly in the ability to forecast the
customers' demand for individual products.  More specifically: trade exactly
what you can sell to your customers.  This task has two aspects if your company
has many branches in different regions: trade what your customers would like
to buy because the product as such is attractive to them and provide a demand
adjusted number of items for each branch or region.

In this paper we deal with the second aspect only: meet the branch distributed
demand for products as closely as possible. The first aspect clearly also
interferes with the total demand for a product over all branches.  Therfore,
we assume that we are given a fix total number of items per product which
should be distributed over the set of branches to meet the the
branch-dependent demand distribution as closely as possible.

Our industry partner is a fashion discounter with more than 1\,000~branches
most of whose products are never replenished, except for the very few
{``}never-out-of-stock{''}-products (NOS products): because of lead times of
around three months, apparel replenishments would be too late anyway.  In most
cases the supplied items per product and apparel size lie in the range between
$1$ and $6$.

The task can be formulated informally as follows: Given historic supply and
sales data for a commodity group, find out some robust information on the
demand distribution over branches in that commodity group that can be used to
optimize or at least to improve the supply distribution over all branches.


We remark that trading fashion has the special feature that also the demand
for different apparel size varies over the branches.  In this article, however,
we focus on the aspect of improving the supply distribution over all branches.
The apparel size distribution problem is subject some other research in
progress.

\subsection{Related work}

Demand forecasting for NOS items is an well-studied topic both in research and
practice.  The literature is overboarding, see, e.g.,
\cite{f_handbook,kok_fisher,sec_forecasting_handbook} for some surveys.  For promotional
items and other items with single, very short life cycles, however, we did not
find any suitable demand forecasting methods.

The literature in revenue management (assortment optimization, inventory
control, dynamic pricing) very often assumes the neglectability of
out-of-stock substitution effects.  This out-of-stock substitution in the
sales data of our partner, however, poses the biggest problem in our case.  In
our real-world application we have no replenishment, small volume deliveries
per branch, lost sales with unknown or even no substitution, sales rates
depending much more on the success of the individual product at the time it
was offered than on the size.  Therefore, estimating the absolute future
demand distribution from historical sales data with no correction for
out-of-stock substitution seems questionable.

Most demand forecasting tools used in practice are provided by specialized
software companies.  Quite a lot of software packages are available, see
\cite{software_survey} for an overview.  Our partner firm has checked several offers in the
past and did -- apart from the NOS segment --- not find any optimization tools
tailored to their needs.

\subsection{Our contribution}

We show that a reasonably-looking attempt to measure the demand distribution
over all branches by measuring for each branch the sales over all products up
to a certain day (to avoid out-of-stock substitution) does not work because of
the high volatility in the sales rates of different products.

The key idea of this work is that estimating something weaker than the
absolute fraction of total demand of a branch will result in stronger
information that is still sufficient to improve on the demand consistency of
the supply of branches.

More specifically, we propose the new Top-Dog-Index (TDI) that can measure the
branch dependent deviation of demand from supply, even for very small sales
amounts or short selling periods.  This yields, in particular, an estimate for
the direction in which the supply was different from demand in the past for
each branch.

On the one hand, the TDI is a rather coarse measurement; on the other hand, we
can show that on our real-world data set it is statistically robust in the
sense that the TDIs of the branches relative to each other are surprisingly
similar on several independent samples from the sales data and their
complements.

To show the value of the information provided by the TDI, we propose a dynamic
optimization procedure that shifts relative supply among branches until the
deviation measured is as small as possible.

Of course, the impact of such an optimizaton procedure has to be evaluated in
practice.  This is subject of future research.


\subsection{Outline of the paper}

In Section~\ref{sec:real-world-problem} we state the real-world
problem we are interested in.  Moreover, we give an abstract problem
formulation.  An obvious approach of determining the demand distribution of
the branches directly from historic sales data is shown to be inappropriate on
our given set of sales data in Section~\ref{sec_real_data_analysis}.  We
propose our new Top-Dog-Index in Section~\ref{sec:tdi}.  We analyze its
statistical robustness and its distinctive character in clustering branches
according to the deviation of the historic ratio between supply and demand. In
Section~\ref{sec:optimization} we describe an heuristic iterative procedure
that uses the information from the Top-Dog-Indices to alter the supply
distribution towards a suitable distribution that more or less matches the
demand distribution over branches. An outlook and a conclusion will be given
in Section~\ref{sec:outlook}.

\section{The real-world problem and an abstract problem formulation}
\label{sec:real-world-problem}

Our industry partner is a fashion discounter with over $1\,000$ branches. 
Products can not be replenished and the number of sold items per product and
branch is rather small. There are no historic sales data for a specific
product available since every product is sold only for one selling period. The
challenge for our industry partner is to determine a suitable total amount of
items of a specific product which should be bought. For this part the
knowledge and experience of the buyers employed by a fashion discounter is
used. We seriously doubt that a software package based on historic sales data
can do better. But there is another task being more accessible for computer
aided forecasting methods. Once the total amount of sellable items of a
specific product is determined, one has to decide how to distribute this total
amount to a set of branches $B$ which differ in their demand. The remaining
part of this paper addresses the latter task.

In the following, we formulate this problem in a more abstract way. Given a set
of branches~$B$, a set of products~$P$, a function~$S(b,p)$ which denotes the
historic supply of product~$p$ for each branch~$b$, and historic sales
transactions from which one can determine how many items of a given
product~$p$ are sold in a given branch~$b$ at a given day of sales~$d$.  The
target is to estimate a demand $\eta(b, \tilde{p})$ for a future product
$\tilde{p}\notin P$ in a given branch~$b$, where we can use $\sum_{b\in
  B}\eta(b, \tilde{p})=1$ as normalization. This estimation $\eta(b,
\tilde{p})$ should be useable as a good advice for a supply $S(b,\tilde{p})$.
No further information, e.g., on a stochastic model for the purchaser
behavior, is available.

\section{Some real-data analysis evaluating an obvious approach}
\label{sec_real_data_analysis}

The most obvious approach to determine a demand distribution over branches is
to count the sold items per branch and divide by the total number of sold
items.  Here we have some freedom to choose the day of the sale where we
measure these magnitudes.  We have to balance two competing influences.  An
early measurement may provide numbers of sale which are statistically too
small for a good estimate.  On the other hand on a late day of sales there
might be too much unsatisfied demand to estimate the demand since no
replenishment is possible in our application.

The business strategy of our partner implies to cut prices until all items are
sold.  So, a very late measurement would only estimate the supply instead of
the demand.  As there is no expert knowledge to decide which is the
\textit{optimal} day of sales to measure the sales and estimate the branch
dependent demand distribution we have adapted a statistical test to measure
the significance of the demand distributions obtained for each possible day of
counting the sold items. Given a data set~$D$, a day of sales $d$ let
$\phi_{b,d}(D)$ be the estimated demand for branch $b$ determined using the
amounts of sold items up to day $d$ as described above.

We normalize the values $\phi_{b,d}(D)$ so that we have $\sum\limits_{b\in
  B}\phi_{b,d}(D)=1$ for each day of sales $d$, where $B$ is the set of
branches. A common statistical method to analyze the reliability of a
prediction based on some data universe~$D$ is to randomly partition $D$ into
two nearly equally sized disjoint samples~$D_1$ and~$D_2$ with $D_1 \dot\cup D_2
= D$ and to compare the prediction based on~$D_1$ with the prediction based
on~$D_2$. If the two predictions differ substantially than the used prediction
method is obviously not very trustworthy or statistically speaking not very
robust.

In the following part of this section we analyze the robustness of the
prediction $\phi_{b,d}(D)$ for every possible sales day, meaning that even
an \textit{optimal} sales day for the measurement does not provide a
prediction being good enough for our purpose. To measure exactly by how much
two predictions $\phi_{\cdot,d}(D_1)$ and $\phi_{\cdot,d}(D_2)$ differ we
introduce the following:

\begin{definition}
  For a given sales day~$d$ and two samples~$D_1$ and~$D_2$ we define the
  discrepancy $\delta_d$ as
  \begin{equation}
    \delta_d(D_1,D_2):=\sum\limits_{b\in B}\left|\phi_{b,d}(D_1)-\phi_{b,d}(D_2)\right|.
  \end{equation}
\end{definition}

\begin{figure}[htbp]
  \begin{center}
    \includegraphics[width=6.75cm]{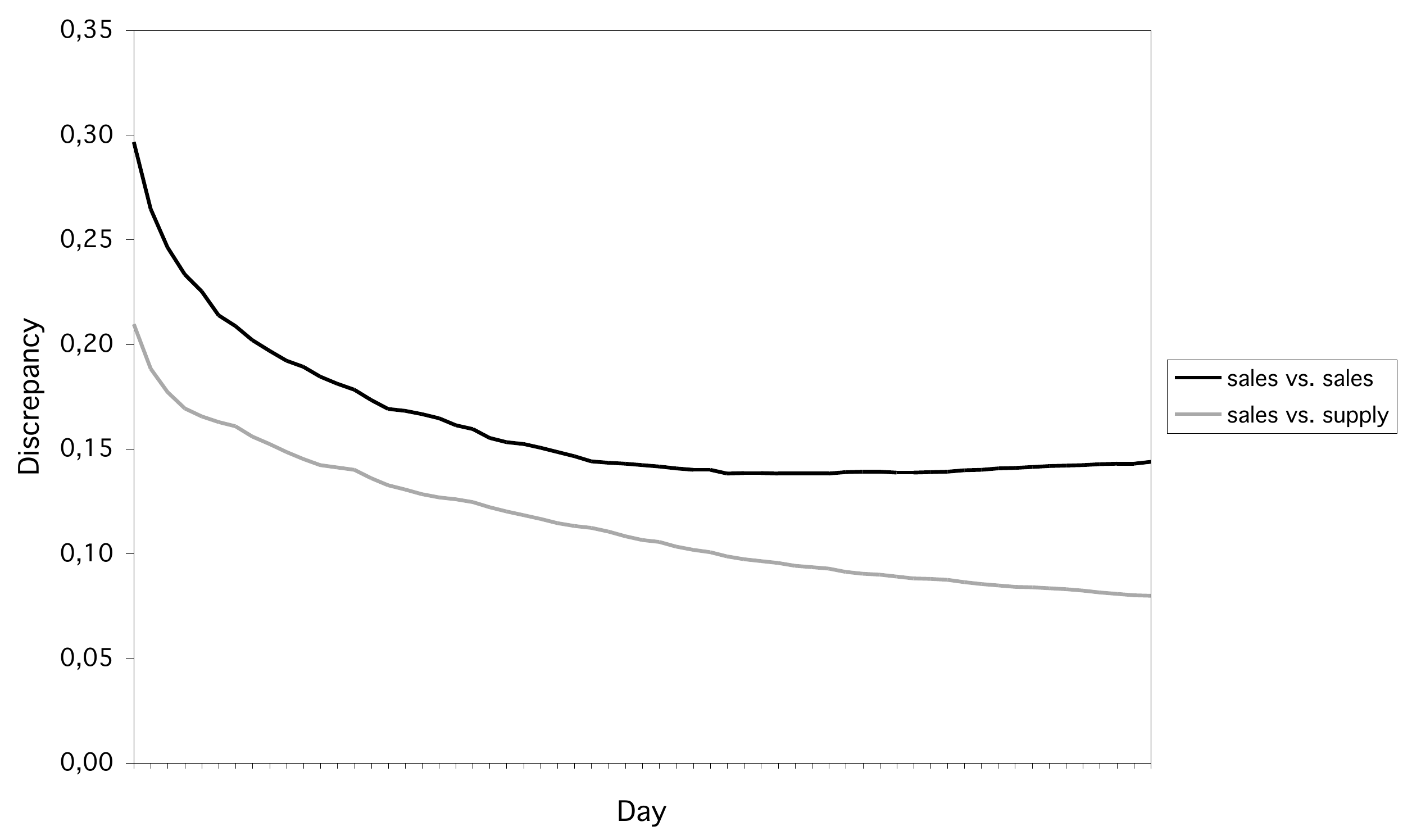}  
  \end{center}
  \caption{Discrepancy for the first 60 days.}
  \label{fig:discrepancy-result}
\end{figure}

Similarly we define a discrepancy between supply and demand. We compare both discrepancies 
in Figure \ref{fig:discrepancy-result}. The result: there is no measuring day for which the discrepancy between two
samples is smaller than the discrepancy between a sample and the supply.  In
other words, if we consider the discrepancy between supply and demand as a
measure for the inconsistency of the supply distribution with the demand
distribution, then either the supply is not significantly inconsistent with
demand (i.e., we should better change nothing) or the measurements on the
various samples are significantly different (i.e., nothing can be learned
about how to correct the supply distribution).
%

\begin{figure}[htbp]
  \begin{center}
    \includegraphics[width=6.75cm]{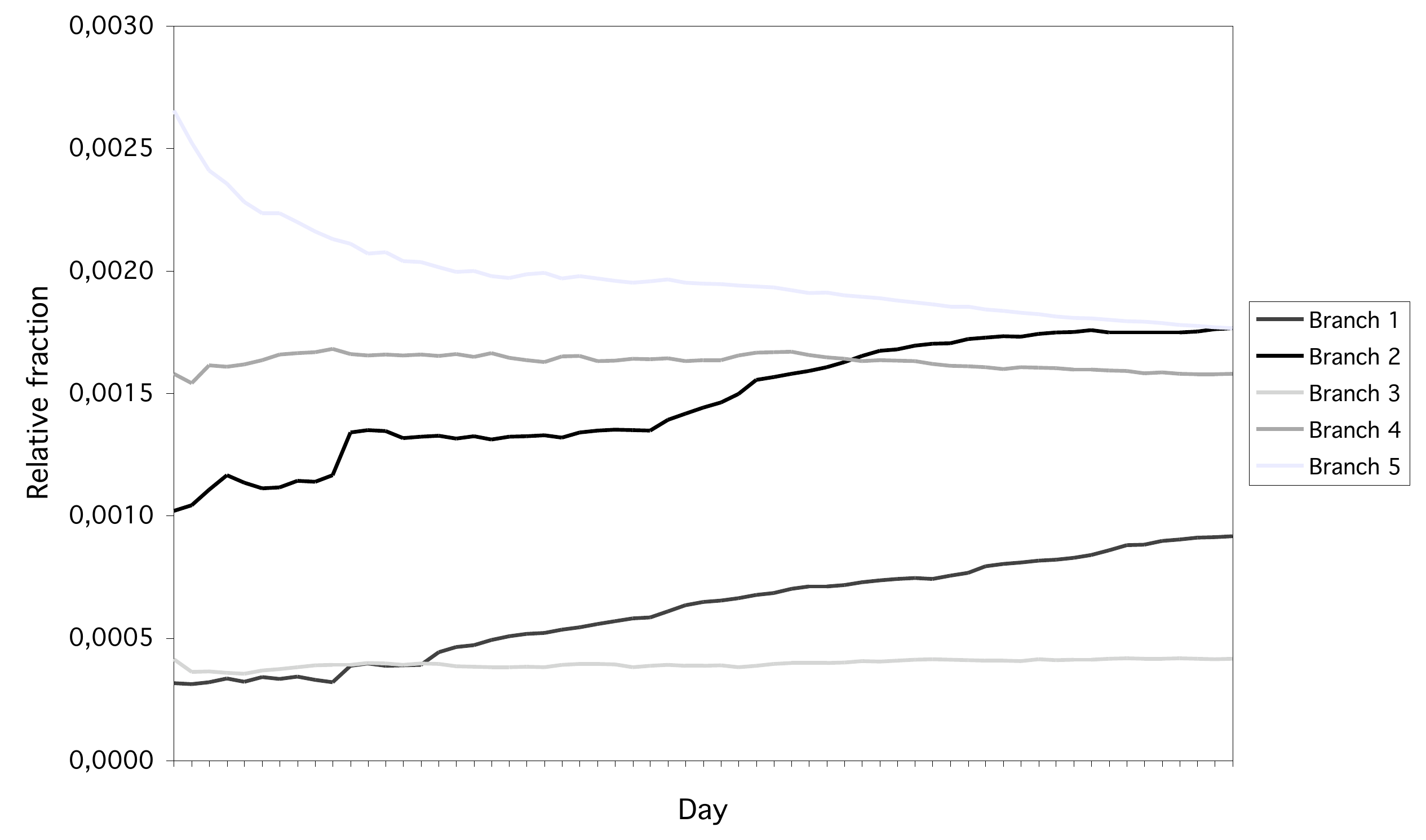}
  \end{center}
  \caption{Prediction $\mathbf{\phi_{b,d}(D)}$ over time.}
  \label{fig:verlauf}
\end{figure}

An explanation why this obvious approach does not work well in
our case is due to the small sale numbers and the interference of the demand
of a branch with product attractivity and price cutting strategies. In Figure \ref{fig:verlauf} we depict the change of prediction $\mathbf{\phi_{b,d}(D)}$ over time for five characteristic but arbitrary branches

We would like to remark that one of the authors currently advises two diploma
theses which check some common parametric models for demand forecasting on
historic sales data from literature.  None of them gives significant
information of the demand distribution over branches of our data set because
the data does not exhibit any similarity to the parametric distributions
coming from economic theory and the like.  This may be due to the fact that
the contaminating effects of promotion, mark-downs, openings/closings of
competing stores prohibit a causal model for the demand.  We do not claim that
the assumptions of parametric demand models never hold, but in our application
they are most certainly not met.

\section{The Top-Dog-Index (TDI)}
\label{sec:tdi}

In the previous section we learned that in our application we cannot utilize
the most obvious approach of looking at the sales distribution over the
different branches on an arbitrary but fixed day of the selling period of each
individual product. Since there is also no indication that any of the common
parametric models for the demand estimation directly from sales data fit in
our application we make no assumptions on a specific stochastic distribution
of the purchaser behavior.

Our new idea dismisses the desire to estimate an absolute percental demand
distribution for the branches.  Instead we develop an index measuring the
relative success of a branch in the competition of all branches that can be
estimated from historic sales data in a stable way.

To motivate our distribution free measurement we consider the following
thought experiment. For a given branch $b$ and given product $p$ let
$\theta_b(p)$ denote the stock-out-day. Let us assume that we have
$\theta_b(p)=\theta_{b'}(p)$ for all products $p$ and all pairs of branches
$b$, $b'$.  In this situation one could certainly say that the
branch-dependent demand is perfectly matched by the supply. Our measure tries
to quantify the variation of the described ideal situation.

Therefore, we sort for each product $p$ the stock-out-days $\theta_b(p)$ in
increasing order. If for a fixed product $p$ a branch $b$ is among the best
third according to this list it gets a \textit{winning point} for $p$. If it
is among the last third it is assigned a losing point for $p$.  With $B_p$
being the set of branches which trade product $p$ and $P$ being the set of the
products traded by the company we can define more precisely:

\begin{definition}
  Let $b$ be a branch. The \textit{Top-Dog-Count} is defined as $W(b):=$
  \begin{equation}
     \left|\left\{p\in P\,\,\Big|\,\, \frac{1}{3}|B_p|\ge|\{b'\in B_p\mid \theta_{b'}(p)\le\theta_b(p)\}|\right\}\right|
  \end{equation}
  and the \textit{Flop-Dog-Count} is defined as $L(b):=$
  \begin{equation}
     \left|\left\{p\in P\,\,\Big|\,\, \frac{1}{3}|B_p|\ge|\{b'\in B_p\mid \theta_{b'}(p)\ge\theta_b(p)\}|\right\}\right|.
  \end{equation}
  For a fix dampening parameter $C>0$ let 
  \begin{equation}
    TDI(b):=\frac{W(b)+C}{L(b)+C}
  \end{equation}
  be the \textit{Top-Dog-Index (TDI)} of branch $b$.
\end{definition}

If the TDI of a branch $b$ is significantly large compared to the TDIs of the
other branches then we claim that branch $b$ was undersupplied in the past.
Similarly, if the TDI of branch $b$ is significantly small compared to the
TDIs of the other branches then we claim that branch $b$ was oversupplied in
the past. We give an heuristic optimization procedure past on this information
in the section. The effect of the dampening parameter $C$ is on the one hand
that the TDI is well defined since division by zero is circumvented. On the
other hand, and more important, the influence of small Top-Dog- or
Flop-Dog-Counts, which are statistically unstable, is leveled to a decreased
importance.

\subsection{Statistical significance of the TDI}

Similarly as in Section \ref{sec_real_data_analysis} we want to analyze the
significance of the proposed Top-Dog-Index on some real sales data. 
Instead of two data sets $D_1$ and $D_2$ we use seven such samples~$D_i$.
Therefore we assign to each different product $p\in P$ a equi-distributed
random number $r_p\in\{1,2,3,4\}$. The samples~$D_i$ are composed as summarized
in Table \ref{tab:test_sets}.

\begin{table}[htbp]
  \begin{center}
    \begin{tabular}{l}
      \hline
      $D_1:=\big\{p|\in P\,\,\big|\,\, r_p\in\{1,2\}\big\}$\\
      $D_2:=\big\{p|\in P\,\,\big|\,\, r_p\in\{3,4\}\big\}$\\
      $D_3:=\big\{p|\in P\,\,\big|\,\, r_p\in\{1,3\}\big\}$\\
      $D_4:=\big\{p|\in P\,\,\big|\,\, r_p\in\{2,4\}\big\}$\\
      $D_5:=\big\{p|\in P\,\,\big|\,\, r_p\in\{3\}\big\}$\\
      $D_6:=\big\{p|\in P\,\,\big|\,\, r_p\in\{1,2,4\}\big\}$\\
      $D_7:=\big\{p|\in P\,\,\big|\,\, r_p\in\{1,2,3,4\}\big\}$\\
      \hline
    \end{tabular}
  \end{center}
  \caption{Assignment of test sets.}
  \label{tab:test_sets}
\end{table}

For the interpretation we remark that the pairs $(D_1,D_2)$, $(D_3,D_4)$, and
$(D_5,D_6)$ are complementary. The whole data population is denoted by~$D_7$
and equals~$P$. We use $TDI(b,D_i)$ as an abbreviation of $TDI(b)$ where $P$
is replaced by $D_i$.

Since the Top-Dog-Index is designed as a non-quantitative index we have to use
another statistical test to assure ourselves that it gives some significant
information. We find it convincing to regard the Top-Dog-Index as significant
and robust whenever we have
\begin{equation}
  \frac{TDI(b,D_i)}{TDI(b,D_j)}\approx\frac{TDI(b',D_i)}{TDI(b',D_j)} 
  \label{eq:tdi}
\end{equation}
for each pair of branches $b,b'$ and each pair of samples $D_i,D_j$. In
words we claim that the Top-Dog-Index is a relative index which is independent
of the underlying sample if we consider a fixed universe~$D_7$. 

\begin{figure}[!htbp]
  \begin{center}
    \includegraphics[width=6.75cm]{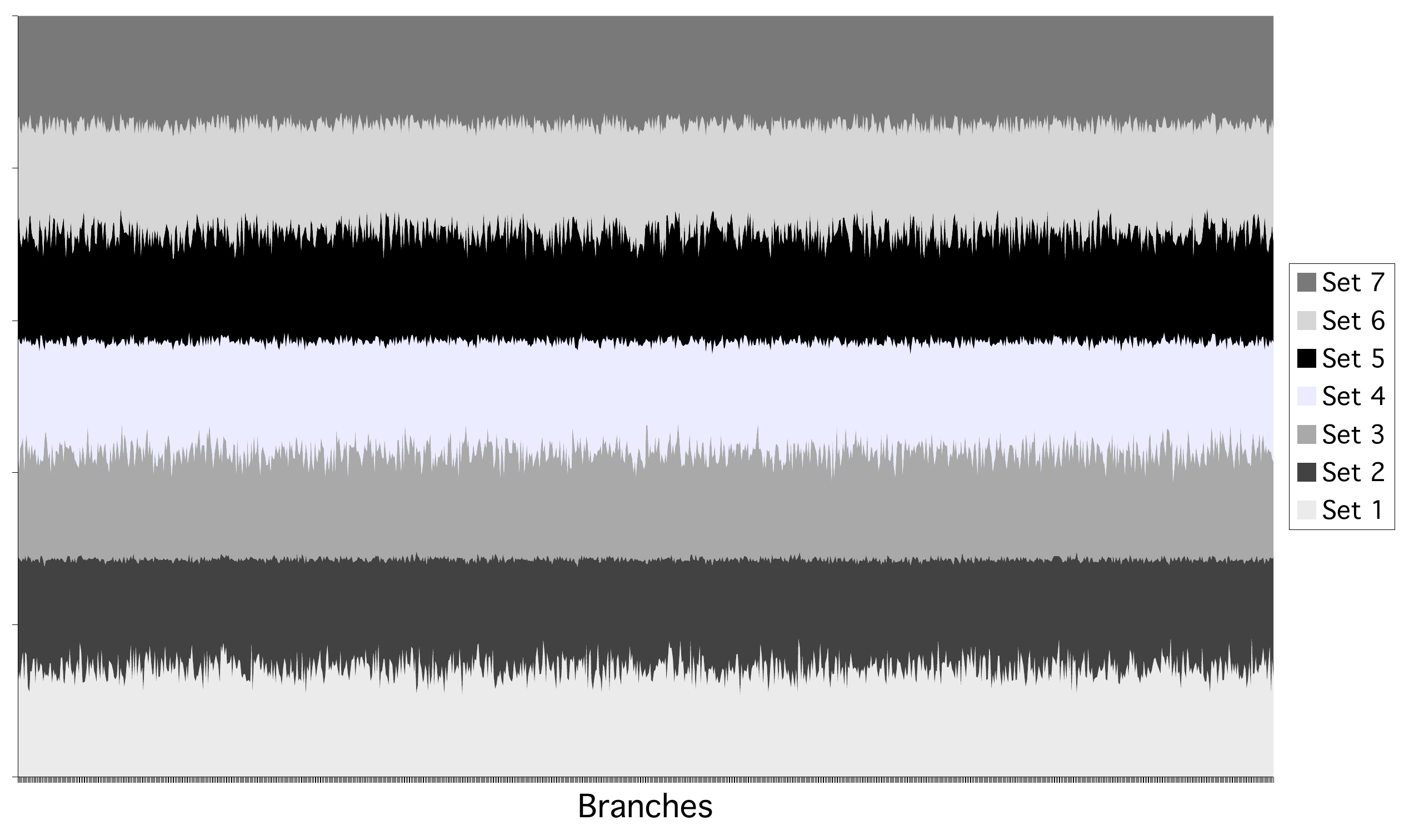}
  \end{center}
  \caption{Relative distribution of the Top-Dog-Index on different data
    samples and branches.}
  \label{fig:tdi-signifigance}
\end{figure}

Our first aim is to provide evidence that the $TDI(b)$ values are robust
measurements.  There is a nice way to look at equation (\ref{eq:tdi})
graphically. For each branch $b$ let us plot a column of the the relative
values $\frac{TDI(b,D_i)}{\sum\limits_j TDI(b,D_j)}$ for all $i$. The result
for our data set is plotted in Figure~\ref{fig:tdi-signifigance}.

\begin{figure}[htbp]
  \begin{center}
    \includegraphics[width=6.75cm]{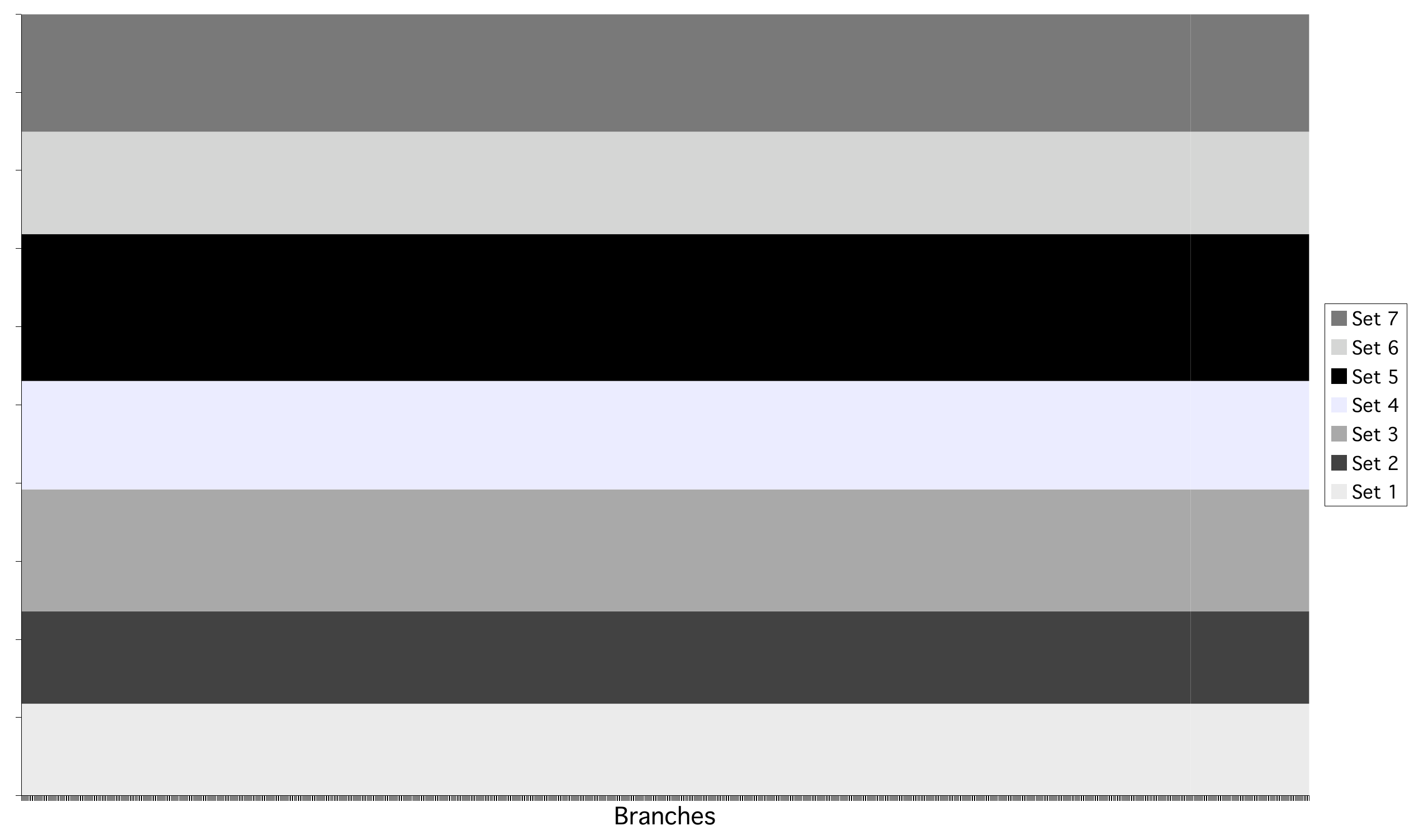}
  \end{center}
  \caption{Relative distribution of deterministic numbers.}
  \label{fig:deterministic}
\end{figure}

\begin{figure}[htbp]
  \begin{center}
    \includegraphics[width=6.75cm]{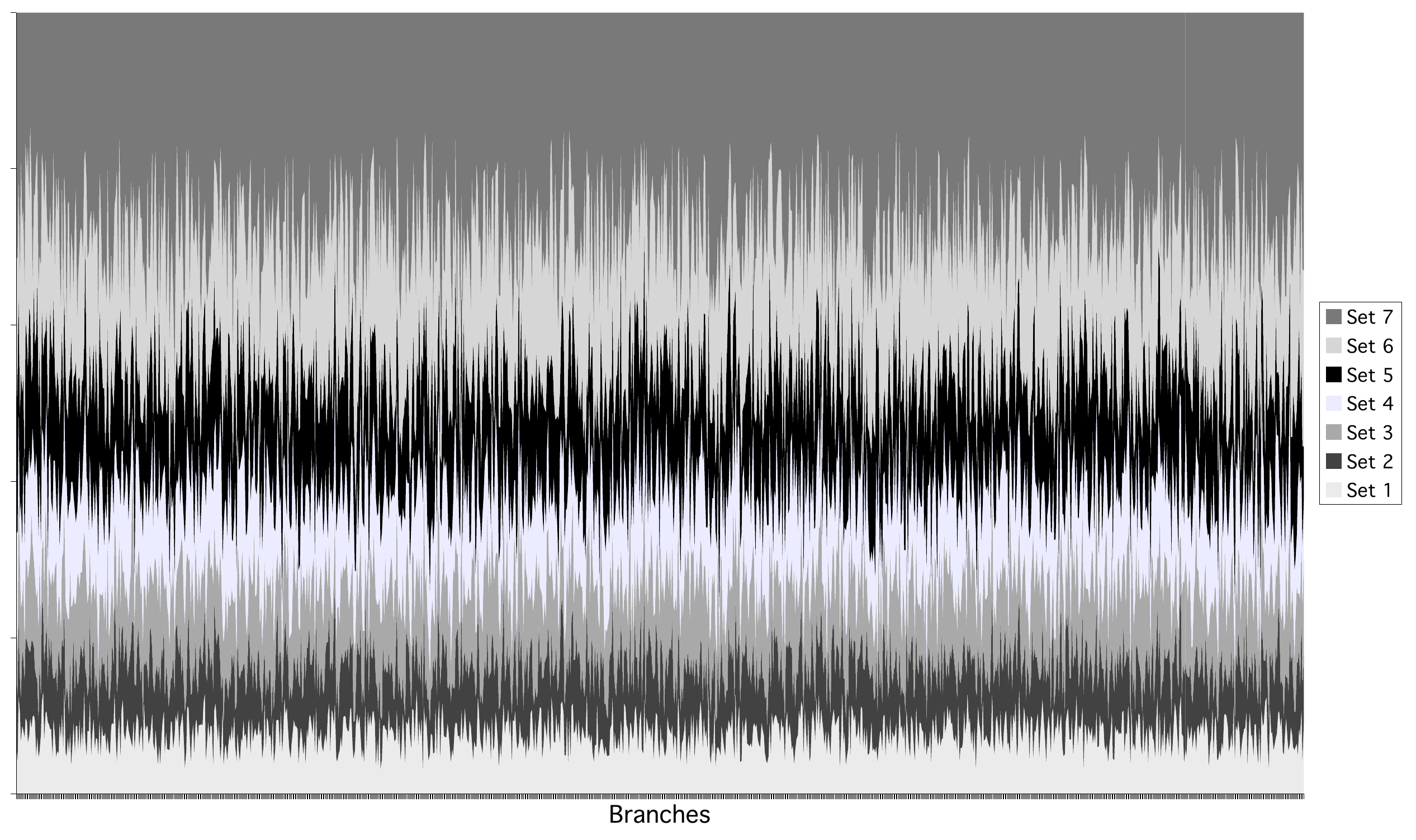}
  \end{center}
  \caption{Relative distribution of in $\mathbf{[0.5,1.5]}$ equi-distributed random variables.}
  \label{fig:random_2}
\end{figure}

To get the correct picture in the interpretation of the plot of Figure
\ref{fig:tdi-signifigance} we compare it to the extreme cases of deterministic
numbers (i.e., $\frac{TDI(b,D_i)}{TDI(b,D_j)}= c_{ij} = c$ for all $i$ and
$j$), see Figure \ref{fig:deterministic}, and random numbers, see Figure
\ref{fig:random_2}.

As a matter of fact, the regions of same color in the plot of the relative
distribution of deterministic numbers in Figure \ref{fig:deterministic} are
formed by perfect rectangles, which are not forced in general to have equal
height.

\vspace*{-0mm}

As an example for a \textit{random plot} we depict in Figure
\ref{fig:random_2} the relative distribution of random numbers being
equi-distributed in the interval $[0.5,1.5]$.

\begin{figure}[htbp]
  \begin{center}
    \includegraphics[width=6.75cm]{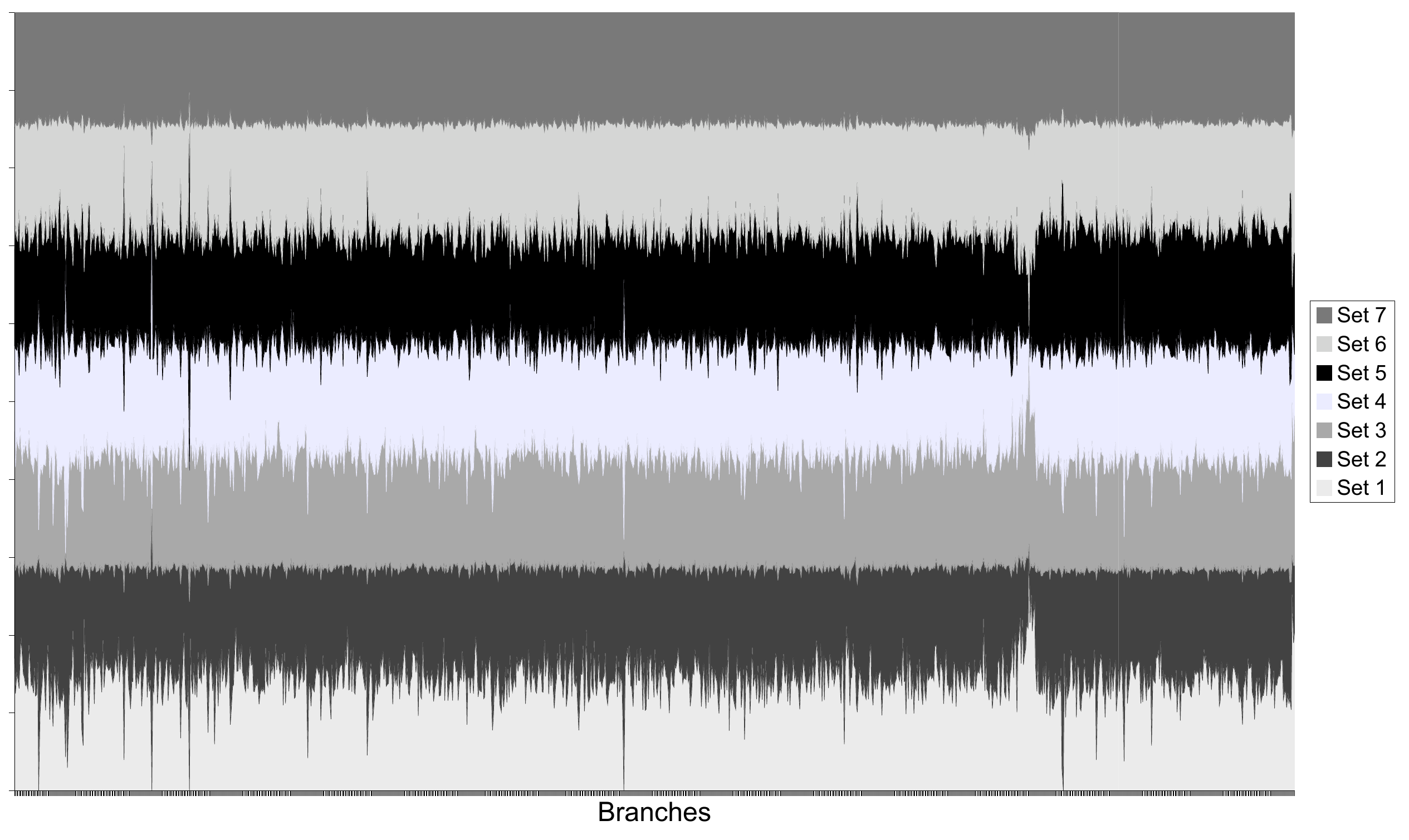}
  \end{center}
  \caption{Relative distribution of $\mathbf{\phi_{\cdot,5}}$.}
  \label{fig:day_5}
\end{figure}

In the plots of Figure~\ref{fig:tdi-signifigance}, \ref{fig:deterministic},
and~\ref{fig:random_2} we can see that that the TDI on the given data set
behaves more like a perfect deterministic estimation than a random number
distribution.  (Ideally, one should now quantify how large the probability is
to obtain a TDI chart as in Figure~\ref{fig:tdi-signifigance} by a random
measurement.) So there is empirical evidence that the TDI gives some stable
information. As a comparison of the TDI and the method described in Section
\ref{sec_real_data_analysis} we depict the corresponding relativ distribution
for measuring day~$5$ in Figure \ref{fig:day_5}.  Although a measurement on
this day was the best we could find, it still produces more severe outliers
than the TDI measurement.

\vspace*{-0mm}


Now the question remains whether this information is enough to cluster
branches into oversupplied and undersupplied ones. More directly: is the
distinctive character of the TDI strong enough? We consider this question in
the next subsection. How the TDI information can be used to iteratively
improve the branch dependent ratio between supply and demand will be the topic
of Section~\ref{sec:optimization}.

\subsection{The distinctive character of the TDI}

If one forces the values of the TDIs to be contained in an interval of small
length, then clearly a plot of the relative distributions would look like the
plot of Figure \ref{fig:deterministic}. As an thought experiment just 
imagine how Figure \ref{fig:random_2} would look like, if we would use random 
numbers being equi-distributed in the interval $[0.9,1.1]$ instead of being
equi-distributed in the interval $[0.5,1.5]$

\begin{figure}[htbp]
  \begin{center}
    \includegraphics[width=6.75cm]{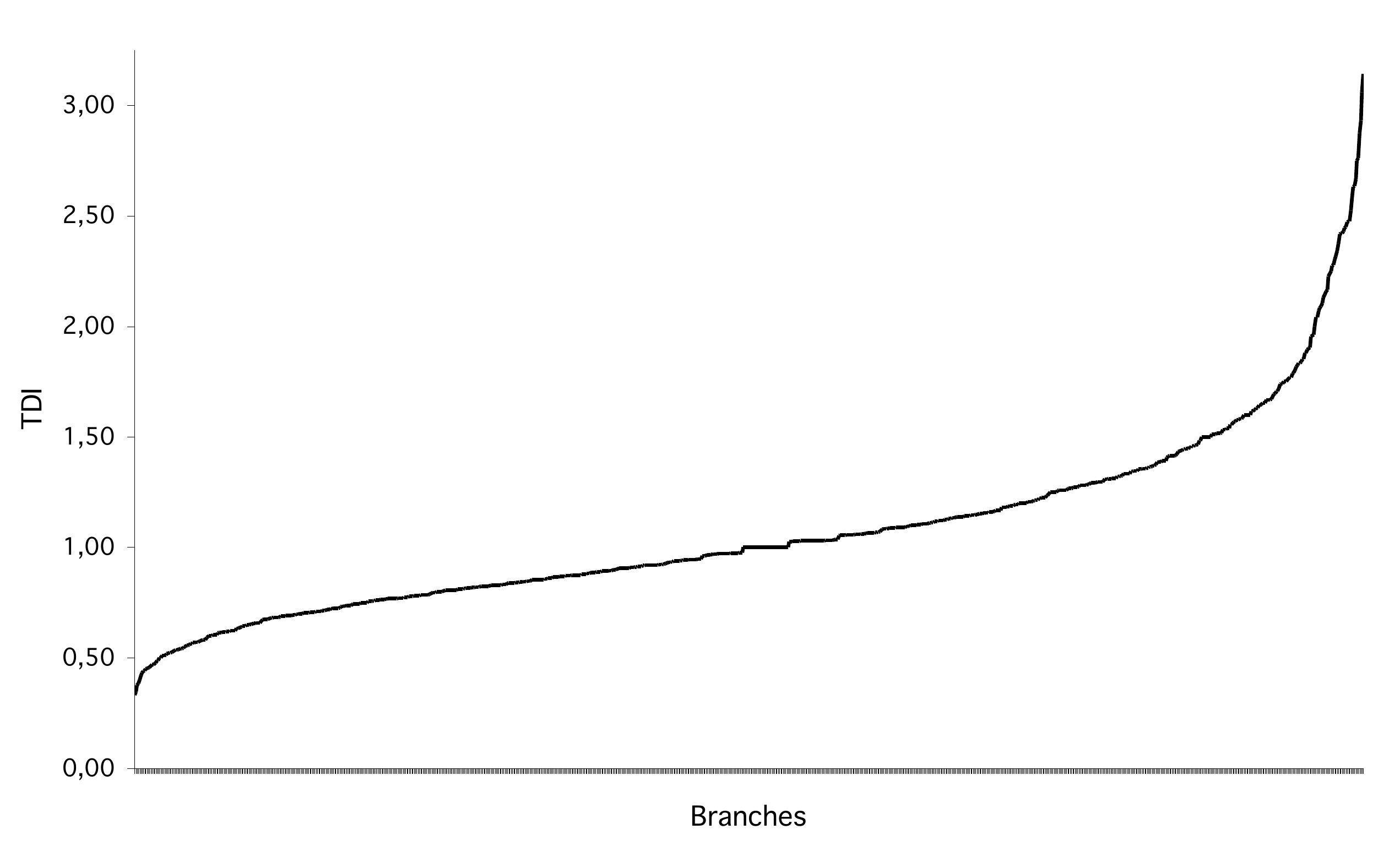}
  \end{center}
  \caption{Occurring TDIs.}
  \label{fig:occuring}
\end{figure}

Forcing the possible values of the TDIs in an interval of small length is
feasible by choosing a sufficiently large dampening parameter $C$. So this
parameter has to be chosen with care. We remind ourselves that we would like
to use the TDIs to cluster branches. Therefore the TDIs should vary over a not
to small range of values to have a good distinctive character.  Clearly by
using the TDI we can only detect possible improvements if the supply versus
demand ratio actually inadequate in a certain level. In Figure
\ref{fig:occuring} we have plotted the occurring TDIs of our data set to
demonstrate there is indeed some variation of values in our data set, no
matter which sample we consider (let alone the data universe).  As one can see
the TDIs vary widely enough to distinguish between historically under- and
oversupplied branches.

\section{The heuristic supply optimization procedure based on the TDI}
\label{sec:optimization}

So far we have developed and statistically stable index capturing the
deviation of supply from demand for each branch. Now we have to specify how we
can use this information to improve the branch dependent ratio between supply
and demand.

Let $S(b)$ be the historic supply of branch $b$ being normalized so that we
have $\sum\limits_{b\in B} S(b)=1$. Our aim is to estimate supplies
$\tilde{S}(b)$, also fullfilling $\sum\limits_{b\in B} \tilde{S}(b)=1$, which
are more appropriate concerning the satisfaction of demand by using the TDI
information.

Therefore let us partition the interval $(0,\infty)$ of the positive real
numbers into a given number of $l$ appropriate chosen intervals
$\mathcal{I}_j$. Further we need $l$ appropriately chosen increment numbers
$\Delta_j$. Our proposed update formula for the estimated branch dependent
demand is given by
\begin{equation}
  \tilde{S}(b)= \frac{S(b)+\Delta_{j(b)}}{\sum_{b' \in B} S(b') + \Delta_{j(b')}}
\end{equation}
for all branches $b$, where $j(b)$ is the unique index with
$TDI(b)\in\mathcal{I}_{j(b)}$.

We do not claim that the $\tilde{S}(b)$ are a good estimation for the demand
of all branches. Our claim is that they approach a good estimation of the
branch dependent demand if one iterates the described procedure over several
rounds and carefully chooses the increment numbers $\Delta_j$, which may vary
over the time.

Once you have a new proposal $\tilde{S}(b)$ of the relative supply for each
branch $b$, one only has to fit it into an integer valued supply for each new
product $p'$.  Given the problem of apparel size assortment and pre-packing,
this is easier said than done and is subject of further studies.

In contrast to the other sections here we are somewhat imprecise and there is
a lot of freedom, e.g., how to choose the intervals $\mathcal{I}_j$ and
increment numbers $\Delta_j$. That is for several reasons. On the one hand
that is exactly the point where some expert form the business should calibrate
the parameters to specific data of the company. One the other hand there are
quite a lot of possibilities how to do it in detail. Their analysis will be a
topic of future research. For the practical application we account rather
simple than sophisticated variants in the first step.

\section{Conclusion and outlook}
\label{sec:outlook}

We have introduced the new Top-Dog-Index which is capable to cluster branches
of a retail company into oversupplied and undersupplied branches at a
statistically robust niveau level where more direct methods fail. The
robustnest of this method is documented by some statistical tests 
based on real-world data. 


We have also documented that the distinctive character of the proposed TDI is
significant for our application: for the first time we can gain information
about the demand distribution of branches from historic sales data on only few
products with volatile success in sales rates and with unknown stock-out
substitution effects, and this information does not depend too much on 
the sample of the sales data universe out of which the TDI is computed.

For the dynamic optimization of the supply distribution among branches, some
fine tuning of parameters is needed; for a real-world implementation these
details have to be fixed. This, together with a field study of the impacts of
an improved supply distribution are research in progress.

\nocite{wop,kouris_a}
\bibliographystyle{latex8}
\bibliography{paper_rio2}

\end{document}